\documentclass[twocolumn,showpacs,preprintnumbers,amsmath,amssymb,nofootinbib]{revtex4}

\usepackage{dcolumn}

\usepackage{amsmath,amsfonts,bbm, graphicx,color}
\def\>{\rangle}
\def\<{\langle}
\def\E{ {\cal E} }

\def\H{ {\cal H} }

\def\M{ {\cal M} }

\def\U{ {\cal U} }

\def\B{ {\cal B} }

\def\F{ {\cal F} }
\def\U{ {\cal U} }
\def\K{ {\cal K} }
\def\R{ {\cal R} }

\def\I{ \mathbbm{1} }
\def\Tr{ \mbox{Tr} }

\def\>{\rangle}
\def\<{\langle}

\def \be{\begin{equation}}
\def \ee{\end{equation}}
\def \beq{\begin{equation}}
\def \eeq{\end{equation}}
\def \bea{\begin{eqnarray}}
\def \eea{\end{eqnarray}}

\newtheorem{theorem}{Theorem}[section]

\begin{document}

\title{The WAY theorem and the quantum resource theory of asymmetry} %
\author{Mehdi Ahmadi}
\author{David Jennings}
\author{Terry Rudolph}
\affiliation{Controlled Quantum Dynamics Theory Group, Level 12, EEE, Imperial College London, London SW7 2AZ,
United Kingdom}%
\begin{abstract}
The WAY theorem establishes an important constraint that conservation laws impose on quantum mechanical measurements. We formulate the WAY theorem in the broader context of resource theories, where one is constrained to a subset of quantum mechanical operations described by a symmetry group. Establishing connections with the theory of quantum state discrimination we obtain optimal unitaries describing the measurement of arbitrary observables, explain how prior information can permit perfect measurements that circumvent the WAY constraint, and provide a framework that establishes a natural ordering on measurement apparatuses through a decomposition into asymmetry and charge subsystems.
\end{abstract}
\date{\today}

\maketitle

\section{Introduction}

The issue of measurement in quantum theory is an old and rich topic, dating back to the foundations of the theory itself. The traditional account tells us that a quantum system $S$ has a set of states described by a Hilbert space $\H$, while an observable $L_S$ of the system is represented as a Hermitian operator on $\H$, whose eigenvalues are the possible values of $L_S$ that can occur in an experiment. A measurement of $L_S$ is described by a set of projection operators (or more generally as a POVM) by which the quantum state is projected onto an eigenstate of $L_S$, corresponding to each particular measurement outcome. However, a measurement of $L_S$ can be described in two distinct ways, firstly in terms of the information acquisition in which our knowledge about a given quantum state $\rho$ is abruptly updated, or secondly by describing the situation from the outside, as a dynamical physical process in which the system $S$ couples unitarily to some quantum mechanical measuring device $A$. This measuring device itself must possess an appropriate `pointer' observable $Z_A$ that serves to `record' the particular value of $L_S$.

The prototypical model of a sharp measurement as a unitary process is the von Neumann-L\"{u}ders measurement \cite{busch} in which the apparatus system is initialized in some default state $|\varphi_0\> \in \H_A$, and then dynamically coupled to $S$ under some unitary $V$ on $\H_S\otimes \H_A$. The observable $Z_A$ is assumed to have a sufficiently large, non-degenerate spectrum to describe all possible measurement values of $L_S$, and the von Neumann-L\"uders measurement is required to be perfectly accurate in that $V$ sends $|e_i\>\otimes |\varphi_0\> \rightarrow |e_i\>\otimes |z_i\>$ for all $i$, where $\{|z_i\>\} $ is the eigenbasis of $Z_A$. The von Neumann-L\"{u}ders measurement, denoted $(\H_A,Z_A, |\varphi_0\>, V)$, has certain distinct characteristics. In particular, it describes a sharp measurement of $L_S$ in the sense that we obtain a single value for the observable, and is also \textit{repeatable} meaning that if $S$ is prepared in an eigenstate $|e_k\>$ of $L_S$, then the measurement process leaves $S$ in this same eigenstate.

In 1952 Wigner provided analysis that showed that in the presence of a conservation law it is impossible to perform an ideal measurement of an observable $L_S$ that does not commute with the conserved quantity \cite{wigner}. Specifically, Wigner showed that if one has an additive conservation law of some quantity $N_{\mbox{\tiny tot}}= N_S\otimes \I + \I \otimes N_A$ over the composite system (such as angular momentum or baryon number), and an observable $L_S$ for which $[L_S, N_S] \ne 0 $, then there cannot exist a von Neumann-L\"uders measurement that respects the conservation law with $[V,N_{\mbox{\tiny tot}}]=0$. Wigner demonstrated, however, that an \emph{approximate} measurement of $L_S$ can be performed, with the error decreasing as a function of the size of the apparatus system. This result was later formalized in the work of Araki and Yanase \cite{araki, Yanase}, where in particular \cite{Yanase} highlighted the necessity that the pointer observable $Z_A$ should commute with the conserved quantity for the apparatus - the \textit{Yanase condition}. The requirement that $[Z_A, N_A]=0$ is in hindsight essential, since otherwise the issue of the measurability of a non-commuting observable can simply be shifted from the system to the apparatus, for example through a swap unitary, and the central problem has merely been postponed. In his paper, Yanase derived a lower bound for the probability of an unsuccessful measurement that scaled as $\<N_A^2\>^{-2}$, a lower bound also obtained by Ghirardi et. al. \cite{Ghirardi}. This lower bound was later tightened by Ozawa \cite{Ozawa} through an application of his generalized uncertainty relation \cite{OzawaGUP}. He found that the root mean square noise $\epsilon(L_S)$ in the measurement of $L_S$ is lower-bounded as
\be\label{Ozawa inequality}
\epsilon^2(L_S)\geq\frac{|\<[L_S,N_S]\>|^2}{4\sigma(N_S)^2+4\sigma(N_A)^2},
\ee
where $\epsilon (L_S)^2 =\<N(L_S)^2\>$ for the noise operator $N(L_S)=V^{\dag}(\I\otimes Z_A)V-L_S\otimes \I$, and $\sigma (X))$ denotes the variance of an observable $X$ in the initial state of the composite system.

The WAY theorem recently received renewed attention by Loveridge and Busch, who have extended to the WAY theorem to continuous variable scenarios such as the joint measurement of position and momentum  \cite{Leonbuschcontinuous}, and have also shown that \emph{both} repeatability and the Yanase condition must be violated if one is to perform a perfect measurement of $L_S$ \cite{Leonbuschreview}. The net result of all these works is the following form of the WAY theorem
\begin{theorem}
\textbf{(WAY)}.
Let $\M(\H_A, Z_A, |\varphi_0\>, V)$ be a von Neumann-L\"uders measurement for an observable $L_S$ on $S$ with eigenstates $\{|e_i\>\}$, with pointer observable $Z_A$ on $A$ with eigenstates $\{|z_i\>\}$. Let $N_S$ and $N_A$ be bounded observables on Hilbert spaces $\H_S$ and $\H_A$, respectively, such that the unitary $V$ obeys $[V,N_S\otimes \I+\I \otimes N_A]=0$. If $\M$ is repeatable or satisfies the Yanase condition, $[Z_A,N_A]=0$, then $[L_S, N_S]=0$.
\end{theorem}

The aim of this paper is to provide an information-theoretic framework that gives a natural and powerful arena in which to analyse arbitrary measurements in the presence of a conservation law. This arena allows us to determine the ultimate constraints on measurements of a given observable under various criteria, to identify the role of prior information in a measurement, to provide an analysis of measurement apparatuses through a subdivision into asymmetry content and charge content, and to describe the role that these two features play in any approximate measurement scenario.

The structure of this paper is as follows. In the next two sections we review the idea of a resource theory of asymmetry, highlight the key properties of the $U(1)$ case that will be needed for our analysis and describe the connection between this resource theory and the presence of a conservation law. In section \ref{equivalence-Umeas-discrim} we establish an equivalence between von Neumann-L\"uders measurements and quantum state discrimination protocols, which leads us to a proof of the WAY theorem in section \ref{WAY-proof}, and provides a neat framework in which to understand the constraints arising from conservation laws. Section \ref{optimal-approx} describes how one can easily construct unitary models that optimally approximate the measurement of observables in a WAY scenario, and elucidates the role that asymmetric resource states and charge eigenstates play in such measurements. Section \ref{examples} illustrates these concepts through the example of a two-dimensional system, by constructing the optimal measurement of an observable $L_S$ for which $[L_S,N_S] \ne 0$, realized as a simple quantum circuit, while \ref{non-trivial-discrim} analyzes a non-trivial optimization scenario using an infinite dimensional measuring apparatus with bounded asymmetry resources and shows the surprising result that under certain natural criteria the optimal resource states do not coincide with the most asymmetric ones. We conclude and discuss our results in \ref{discussion}.

Our notation follows that found in current quantum information theory, and except for $\H$, $\K, \R$ (denoting Hilbert spaces) and $\B (\H)$ (denoting the set of bounded operators on $\H$) we shall denote superoperators using calligraphic script (such as $\E$ and $\F$), and their corresponding Kraus operators written in corresponding Roman script (so for example we would write $\E(\rho) =\sum_iE_i \rho E_i^\dagger $ ). A computational basis is taken simply to be a distinguished orthonormal basis for the system's Hilbert space, and denoted $|0\>, |1\>, |2\>, \dots, |M\>$.

\section{Resource theory of asymmetry, Quantum state discrimination and the WAY theorem }

A useful and unifying concept in quantum information theory is the idea of a consumable ``resource", and its meaning largely coincides with its use in many other contexts. Intuitively speaking, a resource is anything scarce or hard to obtain and which must be consumed in order to achieve some desired action or task. For example, in order to produce mechanical work one must consume free energy, and in order to teleport a quantum state some entanglement must be consumed.

Every quantum resource theory is defined by a set of restrictions on the type of operations that we can perform. Only certain states can be prepared under such restrictions, and the resource states are simply defined as those states which cannot be prepared under the restriction. For the resource theory of entanglement, we define the class of LOCC operations corresponding to allowing Local Operations and Classical Communications, and which provides an axiomatic definition of entangled states as its resource states. Entanglement, in all its different forms, is simply identified as that which does not increase under LOCC.

When in the possession of a resource state $|\psi\>$, we may ask what other resource states can be obtained from $|\psi\>$ through the allowed operations of the theory. In general the allowed transformations between resource states is highly complex, and we often subdivide the problem and consider either strictly deterministic transformations or more general stochastic transformations. Once the allowed transformations between states have been established we have a notion of one state being more of a resource than another, and we may define various measure functions to quantify the particular resource, as we shall illustrate in the next section.

\subsection{The resource theory of asymmetry}\label{asym-resource-sect}
All conservation laws can be identified with particular symmetries, and symmetry groups. Energy and momentum conservation corresponds to translational symmetry in time and space respectively, while angular momentum conservation has associated the rotation group $SO(3)$. The symmetry group $G$ of a conservation law itself defines a class of allowed quantum operations that respect the symmetry action. This in turn defines a resource theory in which asymmetric states (with respect to $G$) are defined as the valuable resources. While a conservation law might define a theory of asymmetric states it turns out that the operations that respect the particular conservation law only form a proper subset of the allowed operations of the asymmetry resource theory, as we shall explain in section \ref{relation-between-cons-resource}.

The presence of a group structure allows us to bring to bear all the usual machinery of representation theory in describing the various constituents of the resource theory. Given a system with associated Hilbert space $\H$, and a unitary representation of the group $U: G \rightarrow \B(\H)$, so that quantum states transform as $\rho \mapsto U(g) \rho U^\dagger (g)$ under the group action. The allowed quantum operations $\E: \B(\H) \rightarrow \B(\H)$ of the theory are the G-covariant operations $\E$ such that $\E(U(g)\rho U^\dagger (g)) = U(g) \E (\rho) U^\dagger (g)$ for all $\rho$ and all $g \in G$ \footnote{This definition is easily extended to the more general case where the quantum map is of the form $\E :\B(\H) \rightarrow \B(\K)$, for Hilbert spaces $\H$ and $\K$ of different dimensions.}. This condition can be expressed compactly as
\be [\E,\U(g)]=0,\,\, \forall g\in G
\ee
where $\U(g)(\cdot)=U(g)(\cdot)U^{\dagger}(g)$ is a superoperator acting on $\B(\H)$ \cite{ImanRob,ImanRobreview}.

The state $\rho$ is then called a \textit{symmetric} state if $[\rho,U(g)]=0$ for all $g \in G$, and \textit{asymmetric} otherwise. Symmetric states are `cheap', being preparable for free within the theory, while asymmetric states, transforming non-trivially under the group action, are the resources. The asymmetric states are often referred to as ``quantum reference frames".

In the same way that entangled states are useful states for quantum information processing tasks \cite{Nielsen&Chuang}, the possession of asymmetric states allow certain tasks otherwise impossible within the constraints of the theory. For example, for $G$ being the rotation group in three dimensions, asymmetric states allow the preservation of quantum information encoded in a particle's spin degree of freedom \cite{SSR-RF,DJ}, or a projective spin measurement along a particular spatial axis \cite{AJR}.

The case $G=U(1)$ will be of particular interest to us, and may be associated to the analysis of phase reference frames, such as with a harmonic oscillator, laser or BEC. This abelian group is generated by an operator $N$ as $U(\theta) = e^{-i \theta N}$, which we may take simply to be a `number' operator with integer eigenvalues. The full Hilbert space then splits up into eigenspaces, or charge sectors, of $N$ for which we write $\H = \oplus_n \H_n$.

The asymmetric states of the theory may be written as $|\psi\>=\sum_n\sqrt{p_n}|\psi_n\>$, where $|\psi_n\>$ is a state lying entirely in the $n$-particle sector $\H_n$ of the operator $N$ on $\H$. The interconversion of $U(1)$-asymmetric states might be achieved deterministically or stochastically through $U(1)$-covariant quantum operations respecting the group symmetry. Given the state $|\psi\>$ we might wish to know whether $|\psi\>$ can be deterministically converted to some other state $|\varphi\>= \sum_n \sqrt{q_n} |\varphi_n \>$, using only $U(1)$-covariant quantum operations alone. For this deterministic case a necessary and sufficient condition is known \cite{GourSpek} and depends only on the two states' distributions over charge sectors, which we write as $\mathbf{p}=(p_1, p_2, \dots)$ and $\mathbf{q} = (q_1, q_2, \dots)$, for simplicity. It can be shown that $|\psi\> \xrightarrow{U(1)-cov} |\varphi\>$ deterministically if and only if $\mathbf{p} = \sum_k w_k T^{(k)}\mathbf{q}$ where $0\le w_k \le 1$ and $\sum_k w_k=1$, and $[T^{(k)}\mathbf{v}]_j = v_{j+k}$. In other words the linear translation map $T^{(k)}$ acts by shifting the components of vectors by $k$ steps to the right for $k$ a non-negative integer or by $|k|$ to the left when $k$ is a negative integer.  For example, with $N=\sum_{n \ge 0} n|n\>\<n|$, the $U(1)$-asymmetric state $\frac{1}{2}(|0\> + |1\> +|2\>+|3\>)$ can be converted deterministically, using only covariant operations, to the state $\frac{1}{\sqrt{2}}(|0\>+|1\>)$ or to the state $\frac{1}{\sqrt{2}}(|1\> + |3\>)$, but cannot be converted to $\frac{1}{\sqrt{2}}(|0\>+|3\>)$ deterministically. However, the latter state may be obtained \emph{stochastically}. Indeed, it turns out that from the uniform superposition state $\frac{1}{\sqrt{N+1}}\sum_{n=0}^N |n\>$ we can stochastically obtain any state of the form $\sum_{n=0}^N a_n |k+n\>$ for any $k \in \mathbb{Z}$ and any $\{a_n\}$ respecting normalization \footnote{See \cite{GourSpek} for a fuller account of the resource interconversions, and conversion rates for stochastic transformations between states.}. Also note that if we take $N$ to be a conserved observable then the above examples show that G-covariant transformations generally violate this conservation law.

Whether we consider deterministic transformations or stochastic transformations, we have that any two states, $|\psi_1\>$ and $|\psi_2\>$, can be related within the theory in one of three ways. It might have that $|\psi_1\> \xrightarrow{U(1)-cov} |\psi_2\>$, meaning they are equally asymmetic and can be \emph{reversibly} interconverted, or it might be that $|\psi_1\> \xrightarrow{U(1)-cov}|\psi_2\>$ only, meaning $|\psi_1\>$ is the state with the greater asymmetry, and can be \emph{irreversibly} converted to $|\psi_2\>$ using covariant operations (or vice versa). Finally it might be the case that no covariant transformation exists between $|\psi_1\>$ and $|\psi_2\>$, meaning the two states are fundamentally \emph{incomparable} within the theory. These relations define a partial order $\prec$ on the space of states where $\rho \prec \sigma$ if and only if $\rho$ can be obtained from $\sigma$ by covariant operations. The order derived from deterministic transformations, $\prec_d$, is a strictly stronger relation than that derived from stochastic transformations, $\prec_s$. More specifically, this means that a pure state $|\psi\>$ defines a \textit{stochastic branch} of pure states $\{ |\varphi\> :|\varphi\> \prec_s |\psi\>\}$, which contains as a proper subset within it the \textit{deterministic branch} of pure states $\{|\varphi\>: |\varphi\> \prec_d |\psi\>\}$.

Any real-valued function that respects the stochastic partial ordering provides us with a measure of asymmetry, and certain particularly natural measures of pure state asymmetry already exist. If we choose $|\varphi\> = \frac{1}{\sqrt{2}} (|0\>+ |1\>)$ as our basic unit of asymmetry (an `asbit' \cite{vanEnk}), and consider conversion rates involving asymptotically many copies of $|\psi\>$, we find (for $\mathbf{p}$ being gapless) that
\begin{eqnarray}
|\psi \>^{\otimes M} \xrightarrow{U(1)-cov} |\varphi\>^{\mathrm{Var}(\psi) M}
\end{eqnarray}
where $\mathrm{Var}(\psi) = 4( \< \psi|N^2 | \psi\> - \<\psi| N |\psi\>^2)$ is four times the variance of $N$ in the state $|\psi\>$.
Another such measure is the ``relative entropy of frameness" \cite{ImanGiladRobREF}, for a pure state $|\psi\>=\sum_n\sqrt{p_n}|n\>$, which turns out to be $H(\mathbf{p})$, the Shannon entropy of the distribution $\{p_n\}$. For a system of dimension $M+1$ (or restricting to states in the subspace of sectors $\H_0 \oplus  \cdots \oplus \H_M$) both these measures attain their maximum value on the uniform superposition state,
\begin{eqnarray}\label{optimalapparatus}
|\Psi\> = \frac{1}{\sqrt{M+1} } ( |0\>+|1\>+\cdots +|M\>).
\end{eqnarray}
In this sense, one can then identify the uniform superposition state (\ref{optimalapparatus}) as the most asymmetric pure state with support entirely in $\H_0\oplus \dots \oplus \H_M$, however subtleties arise when we consider optimizing certain tasks. Often our measure of success of a task is expressed in terms of some probability that involves a potentially complex chain of conditionals, and so, in the absence of task details, it is only possible to pronounce the state $|\Psi\>$ as optimal if we restrict to its deterministic branch. If, however, we enlarge our scope to the full stochastic branch of $|\Psi\>$, being all the pure states of the system, any probabilistic measure of success must include the conversion probabilities in going stochastically from $|\Psi\>$ to some other state, and so it may happen that for a particular task the optimal state differs from $|\Psi\>$. We provide an explicit example of this feature in the context of a WAY scenario (for which $|\Psi\>$ was the state originally considered in \cite{wigner, araki, Yanase}) in section \ref{non-trivial-discrim}.

\subsection{Relation between additive conservation laws and $U(1)$ resource asymmetry}\label{relation-between-cons-resource}

As noted earlier, the constraint of a conservation law is a strictly stronger one than the asymmetry constraint of its associated group. For example, energy conserving operations can only transform states within each individual energy sector, while its associated $U(1)$ resource theory (defined by time evolution $U(t)=\exp(-itH)$) would include $U(1)$-covariant transformations that increase or decrease the total energy. Indeed reversible energy conserving transformations correspond to $U(1)$-\emph{invariant} unitaries $V$ for which $[U(t), V]=0$ for all $t$.

One might expect that the units of energy (or conserved `charge') available for use within a bounded apparatus might play a role in addition to any issue of asymmetry. For example, any addition of units of the conserved quantity in the apparatus allows a greater range of transformations through energy conserving couplings between the system and the apparatus. This physical intuition is made more concrete by considering Stinespring dilations of the allowed quantum operations under the $U(1)$-constraint. In its Schr\"odinger form for G-covariant operations it states that \cite{KeylWerner}
\begin{theorem}
Given a $G$-covariant trace-preserving, completely positive map $\E: \B(\H) \rightarrow \B (\H)$, there exist a dilating system $\K$ carrying a representation of $G$, a $G$-invariant unitary $V$ on $\H\otimes \K$, and a $G$-invariant state $|\varphi\>$ in $\K$ such that $\E(\rho) = \Tr_\K [ V(\rho\otimes |\varphi\>\varphi| ) V^\dagger ]$. Moreover, if $\E = \sum_i \E_i$ for G-covariant CP maps $\{ \E_i \}$, then there exist positive operators $\{F_i\}$ on $\K$ such that $\sum_i F_i = \I$ and $[U_\K (g), F_i ]=0$, with $ \E_i (\rho) = \Tr_\K [(\I \otimes F_i V(\rho \otimes |\varphi\>\<\varphi| )V^\dagger ]$.
\end{theorem}

Applied to the $G=U(1)$ case, we conclude that the set of unitary dynamics on a composite system of $S$ and $\K$ that respect the additive conservation law $N_S \otimes \I + \I \otimes N_\K$ (such as in the WAY-theorem scenario) together with the ability to introduce eigenstates of $N_\K$ coincides with the set of $U(1)$-covariant quantum operations allowed on the system $S$, in which quantum coherence in the eigenbasis of the conserved quantity $N_{\mbox{\tiny tot}}$ constitutes a resource. If the spectrum of $N_\K$ is unbounded then the particular eigenstate $|\varphi\>$ used is largely a matter of choice. 

The conservation law dictates that the total amount of the conserved quantity in the closed system never varies, which holds true also for classical systems, however the generation of asymmetric resources is a more subtle prohibition and in a sense should be viewed as an additional, non-classical constraint\footnote{E.g.\!\! in thermodynamics one can generate energy asymmetry without changing the total internal energy of the system, see \cite{SJ1, SJ2} for a discussion.}. The possession of eigenstates of the conserved quantity allows us to perform many useful tasks under the conservation constraint, however the possession of asymmetric resources greatly extends the set of things that we can do.

\subsection{Equivalence of von Neumann-L\"uders measurements and quantum state discrimination}\label{equivalence-Umeas-discrim}
The WAY theorem can now be cast within the framework of asymmetric resources. We consider the situation of some additively conserved quantity $N_
{\mbox{\tiny tot}}$ over two systems $S$ and $A$, and demand that any unitary $V$ respect the conservation law in that $[V, N_{\mbox{\tiny tot}}]$. We consider some arbitrary observable $L_S$ for the system $S$ and attempt to construct a unitary model for its measurement. We know from the previous section that the conservation law scenario is equivalent to that of a $U(1)$-asymmetry constraint on $S$ alone and view the apparatus $A$ as the dilating system of any covariant map on $S$, in which any covariant operation gives rise, via a Stinespring dilation, to conserving unitary dynamics coupling the system to the apparatus $A=\K$ prepared in an eigenstate of $N_A$ followed by a measurement of an observable $Z_A$ that commutes with $N_A$. We can thus focus solely on the system $S$ and restrict to covariant maps $\F$, safe in the knowledge that they can be obtained from some conserving unitary $V$.

We denote the spectral decomposition of $L_S$ as $L_S=\sum_k l_k |e_k\>\<e_k|$, and for simplicity consider a non-degenerate spectrum. In the absence of any constraints, a projective measurement of $L_S$ is described by the trace-preserving operation $\M = \sum_k \M_k $, with $\M_k (\rho) = |e_k\>\<e_k|\rho |e_k\>\<e_k|$ for all $k$. Our task in the presence of constraints is then to obtain $U(1)$-covariant superoperators $\{\F_k\}$ such that the POVM $\{\F_k \}$ is as close as possible to $\{\M_k \}$, where ``close" must now be given some operational meaning.

To define some measure of performance, we may recast the goal of performing a measurement of $L$ in more information-theoretic terms as the encoding of \emph{classical} information in the eigenbasis of $L_S$, where the classical information is encoded in the label of the eigenstate, $k \mapsto |e_k\>\<e_k|$. A state diagonal in this eigenbasis is prepared, and given by $\rho= \sum_i p_i |e_i\>\<e_i|$, where $\{p_i\}$ describes the distribution of the classical source. In the absence of any constraints, a faithful readout of the signal is always possible through the application of any $\M$, for which (I) $\M_i(|e_k\>\<e_k|) = 0$ for $i \ne k$ and (II) $\sum_i\Tr [\M_i [\rho]]=1$ for any $\rho$. However in the presence of the conservation law, it may be that the ideal $\M$ is not a covariant operation, and so the best we can achieve under the constraints is some \emph{approximate discrimination} of the eigenstates of $L_S$ which will fail to satisfy both (I) and (II). Two natural approximate discrimination protocols are unambiguous discrimination (UD) and maximum likelihood estimation (MLE), with each corresponding to the weakening of one of the two central conditions of perfect state discrimination.

For unambiguous discrimination, we still demand that $\F_i( \rho_k) = 0$ for $i \ne k$, but now allow the possibility that $\sum_i \F_i$ is a trace-decreasing POVM map. The interpretation of the first condition is that in obtaining outcome $i$ of the POVM we are certain that $\rho_i$ must have been prepared - we have discriminated unambiguously - while the second condition of allowing trace-decreasing $\sum_i \F_i$ means that sometimes the protocol may fail entirely and we learn nothing about the state. The full quantum operation must conserve probability, and so is described by a total trace-preserving operation $\F = \sum_i \F_i + \F_*$ for which $\F_i (\rho_k)=0$ for $i \ne k$, and $\Tr[\F_* (\rho)]$ being the probability of failure. The goal of UD is to minimize this probability, or equivalently to maximize the probability $\sum_i p_i \Tr[\F_i (\rho_i)]$ of successfully identifying the prepared state.

While UD is essentially the scenario considered in Wigner's seminal paper, recasting the problem in abstract, information-theoretic terms allows us see that another perfectly natural possibility to consider is that of maximum likelihood estimation. Maximum likelihood estimation decides instead to place the short-fall on the first condition. In other words, we enforce that $\sum_i \F_i$ is trace-preserving, but now allow the possibility of approximate discrimination $\F_i(\rho_k) \approx 0$ for $i \ne k$. The goal of MLE is to maximize the probability $\sum_i p_i \Tr[\F_i (\rho_i)]$ of successfully identifying the prepared state.

In the unconstrained setting and when $p_i>0$ for all $i$, the projective measurement $\M$ in the eigenbasis of $L_S$ is singled out as the optimal measurement proceedure to distinguish the states in the ensemble $\{p_i, |e_i\>\<e_i|\}$ perfectly. However this perspective allows us to do more if we wish, and account for prior information as to what state the system was prepared in initially. For example, one might be limited in the particular operations that we can perform, but knowing that our system was prepared with support only in some subspace means that a faithful measurement of $L_S$ may yet be possible though a quantum operation distinct from $\M$.

In the $U(1)$-constrained scenario associated to the conservation of $N_S$, the measurement of $L_S$ becomes the task of optimally discriminating its eigenstates using only $U(1)$-covariant POVM maps $\F=\{ \F_i\}$. For the case of unambiguous discrimination this amounts to minimizing $\Tr[ \F_* [\rho]]$ for $\rho = \sum_k p_k \rho_k =\sum_{k=1}^D \frac{1}{D} |e_k\>\<e_k|$, where for simplicity we do not assume any prior information as to what eigenstate is being prepared.

\subsection{Proof of the WAY theorem}\label{WAY-proof}
All the necessary pieces are now in place. A unitary model for a measurement of the observable $L_S$ in the presence of an additive conservation law $N_{
\mbox{\tiny tot}} = N_S + N_A$ defines a constrained discrimination protocol of orthogonal states within a $U(1)$-asymmetry theory, and conversely any such constrained discrimination protocol defines a unitary model of some measurement in the presence of a conservation law. Whether such a unitary model is possible, or to what degree an approximate model exists, is then determined by the theory of quantum state discrimination under the constraint of covariance.

At the simplest level within a $G$-asymmetry scenario, a distinction is drawn between symmetric states and asymmetric states; between covariant superoperators and non-covariant superoperators. However, given an asymmetric state $\rho$ one can always obtain a symmetric state $\bar{\rho}$ from it through an averaging over the group $\rho\rightarrow \bar{\rho} = \int dg\, U(g)\rho U^\dagger (g)$, called G-twirling, and in the same way, given some non-covariant operator $\E$ one can obtain a $G$-covariant map through the super-operator G-twirling $\E \rightarrow \bar{\E} := \int dg \, \U(g) \circ \E \circ \U^\dagger (g)$. Both these maps are idempotent and are the projectors onto the set of symmetric states and the set of $G$-covariant maps respectively. Consequently, the minimization of $\Tr[\F_*(\rho)]$ over the set of $U(1)$-covariant quantum operations is equivalent to the minimization of $\Tr[\bar{\F}_*(\rho)]$ with $\{\F_i, \F_* \}$ taken over the full set of quantum operations. However $\Tr[\bar{\F}_*(\rho)] = \Tr[\F_* (\bar{\rho})]$, and in a similar way the discrimination condition (I) can be written $ \F_i (\bar{\rho}_k) = 0 $ for $i \ne k$. In other words, we can reformulate our optimization task to that of an \emph{unconstrained} unambiguous discrimination of the G-twirled ensemble $\{p_i, \bar{\rho}_i \}$.

The implications of this are immediate. A unitary model for the measurement of $L_S$ exists if and only if we can discriminate $\{\bar{\rho}_i\}$ perfectly. This is true if and only if $\{ \bar{\rho}_i\} $ have orthogonal supports. Assuming that $\{ \rho_i\}$ has support on a full eigenbasis of $L_S$, if any of the states in the G-twirled ensemble have rank larger than one, then they must overlap with at least one other state in the ensemble and so a perfect von Neumann-L\"uders measurement is impossible. Thus, such a measurement will exist if and only if all G-twirled states are rank one, in which case we have that $\int dg \, U(g) |e_k\>\<e_k| U^\dagger (g) = |\varphi_k\>\<\varphi_k|$. However pure states are the extremal points of state space and so $|\varphi_k\>\<\varphi_k | = |e_k\>\<e_k| = U(g)|e_k\>\<e_k|U^\dagger(g)$ for all $g$, and so $L_S$ must commute with $N_S$, which completes the proof of the WAY theorem from the resource theory perspective.

\subsection{Optimal von Neumann-L\"uders measurement of non-commuting observables}\label{optimal-approx}

The previous analysis identifies when a perfect von Neumann-L\"uders measurement of an observable $L_S$ can occur in the presence of the conservation law, however the formulation allows us to go beyond simply achieving the projective measurement $\{\M_i\}$. We immediately see that the optimal approximate measurement that respects the conservation law will correspond to the optimal discrimination protocol for the G-twirled ensemble $\{p_k,\bar{\rho}_k \}$. In section (\ref{examples}) we provide explicit examples of such optimal protocols, but before that we describe two ways in which the constraint of a conservation law  on the measurement of a non-commuting observable can be overcome.

\subsubsection{The possession of prior information}
It turns out that perfect measurement of a non-commuting $L_S$ may well be possible in the presence of prior information. If our prior information is such that some $p_i$ are zero then it may occur that the G-twirled states are all mutually orthogonal, despite being mixed states. For this situation a perfect measurement of the observable $L_S$ is possible, despite $L_S$ not commuting with the conserved quantity. Phrased another way, in the presence of a conservation law each observable $L_S$ has a `blurring' scale corresponding to the number of eigenstates of $N_S$ in the expansion of the eigenstates of $L_S$. The blurring extreme occurs for observables with eigenstates being fully unbiased with respect to those of $N_S$, and so G-twirl to maximally mixed states. Prior information for states with coherence in the basis $|e_k\>$ can be handled equally well since the presence of the $U(1)$ constraint implies that the coherent prior information should be decohered in the eigenbasis of $L_S$.

\subsubsection{The possession of asymmetry resource states}

When $L_S$ does not commute with the conserved quantity its G-twirled eigenstates will overlap and only an approximate state discrimination is possible, whether under UD or MLE. Indeed for the extreme case that all its eigenstates G-twirl to the maximally mixed state we find that no discrimination, and hence no perfect von Neumann-L\"uders measurement, is possible at all. However such scenarios are not as final as they might first appear. The key idea is that while we are constrained to performing only certain types of operations it might be that we are initially in possession of valuable resource states, whose presence enable otherwise impossible transformations.

The possession of an additional system $\R$ in an asymmetric state $|\Psi\>$ allows us to better encode the eigenstates of the observable $L_S$ in preparation for the discrimination protocol. Specifically, defining $|g\> :=U(g)|\Psi\>$, we can define a sequence of quantum operations
\begin{eqnarray}
|e_k\> \rightarrow |\Psi\> \otimes |e_k\> \rightarrow \int dg \, |g\>\<g| \otimes U(g)|e_k\>\<e_k|U^\dagger (g),
\end{eqnarray}
which is no longer the maximally mixed state. The non-trivial transformation of the state $|\Psi\>$ under the group provides a quantum reference frame, which allows the (partial) encoding of the state $|e_k\>$ into the \emph{relational} degrees of freedom of the composite $G$-invariant state \cite{steveterryrobQC}. From the perspective of the von Neumann-L\"uders measurement, the asymmetry resource system constitutes a distinct part of the measuring apparatus, and so we in general have that $A = \R \otimes \K$, where $\K$ accounts solely for the \emph{sharp} units of conserved charge required within $A$.

\section{Explicit examples}\label{examples}
We can now illustrate the preceeding ideas with the explicit example of
 obtaining a probabilistic von Neumann-L\"uders model that describes the measurement of an observable $L_S$ of a two-dimensional quantum system, with eignstates $|e_+\> =1/\sqrt{2}(|0\>+|1\>)$ and $|e_-\>= 1/\sqrt{2} (|0\> - |1\>)$ while still respecting a conservation law of the observable $N_S= \sum_{n \ge 0} n |n\>\<n| =|1\>\<1|$. This scenario describes, for example, the situation of a spin-1/2 particle with angular momentum conserved only along the $Z$-direction, or the situation of a photon number state in quantum optics, as in the original considerations of Wigner and Yanase \cite{wigner, Yanase}.

It is readily seen that the action of the group transformation $U(\theta) =\exp(i\theta N)$ $G$-twirls both $|e_+\>$ and $|e_-\>$ to the maximally mixed state $\frac{1}{2} (|0\>\<0| + |1\>\<1|)$. Hence, in the absence of any resource state it is impossible to even approximately perform a von Neumann-L\"uders measurement of the observable $L_S$.
\subsection{Uniform superposition states as asymmetry resources}
What about if we have in our possession some resource asymmetry? For simplicity we consider having a uniform superposition of number states $|\Psi\> = \frac{1}{\sqrt{M+1}} (|0\> + \cdots +|M\>)$, which under the conditions discussed earlier, is a maximally asymmetric state for a resource system $\R$, of dimension $M+1$.

\begin{figure}[t]
\centering
\includegraphics[width=6.5cm, height=4cm]{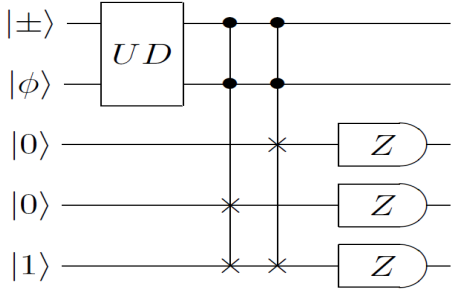}
\caption{Quantum circuit for a von Neumann-L\"uders measurement of the observable $L_S$. }
\label{circuit}
\end{figure}

We find that the states $|\Psi\>\otimes |e_\pm\>$ G-twirl to the mixed states
\begin{eqnarray}
\bar{\rho}_\pm &=& \sigma + \frac{1}{M+1}\sum_{n=1}^{M} |\phi^\pm_n\>\<\phi^\pm_n|
\end{eqnarray}
where we have $\sigma = \frac{1}{2(M+1)}(|0,0\>\<0,0|+ |M,1\>\<M,1|)$, while the remaining eigenstates are given by $|\phi^\pm_n\> = \frac{1}{\sqrt{2}} (|n,0\> \pm |n-1,1\>)$ for $n$ between 1 and $M$. The states $\rho_\pm$ have overlapping support only on the space span$(|0,0\> , |M,1\>)$, whereas they are orthogonal on the rest of the space. As $M$ increases the two states approach orthogonality, and hence become perfectly distinguishable. Thus, in the limit of an infinite reference frame system ($M\rightarrow \infty$) perfect measurement of the observable $L_S$ becomes possible.

The states in the G-twirled ensemble $\{p_\pm, \bar{\rho}_\pm \}$ will always be block diagonal matrices in the conserved quantity basis. The full Hilbert space for the primary system and resource system splits up into the eigensectors of $N_S$ as $\H = \bigoplus_n \H_n$ with $\Pi_n$ being the projector onto sector $\H_n$, and $n$ running from zero to $M+1$. This decomposition simplifies the analysis for obtaining the optimal UD measurement, since it turns out \cite{Raynal} that if $\{\F^{(n)}_k \}$ is the optimal POVM for the UD of the projected ensemble $\{ p_k \Pi_n \bar{\rho}_k \Pi_n \}$ then $\F = \{\F_k \}$ with $\F_k = \sum_n \F^{(n)}_k$ is an optimal POVM for the original ensemble $\{p_k, \bar{\rho}_k \}$.

Since the projection of the states $\bar{\rho}_\pm$ into the subspaces with total number $n=1,2,...,M$ are orthogonal we have that perfect discrimination is possible in each sector $\H_n$ simply through the projective measurement onto the basis $\{|\phi_n^+\>, |\phi_n^-\>\}$. In contrast, the two states $\bar{\rho}_+$ and $\bar{\rho}_-$ are identical when projected onto the one-dimensional sectors $\H_0$ and $\H_{M+1}$, and so all measurements fail to provide any information.

We deduce that the optimal POVM measurement for the twirled ensemble $\{ p_\pm, \bar{\rho}_\pm\}$ is given by $\{\F_+, \F_-, \F_* \}$ where $\F_\pm$ are projection maps given by the rank $M$ projectors $\sum_{n=1}^M |\phi^\pm_n\>\<\phi^\pm_n|$, while $\F_*$ is the rank $2$ projection onto the `bad' sectors $\H_0$ and $\H_{M+1}$, which occurs with probability $\frac{1}{M+1}$.

By the covariant Stinespring theorem, we know that this optimally discriminating POVM can be dilated to a unitary model in which
\begin{eqnarray}
\F_{\pm,*} (\rho) = Tr_\K [(\I\otimes F_{\pm,*})V(\rho \otimes |\varphi\>\<\varphi|)V^\dagger]
\end{eqnarray}
for some unitary $V$ respecting the conservation law, some state $|\varphi\>$ invariant under the group action and POVM elements $F_{\pm,*}$ on $\K$, also invariant under the group action. The elements $F_{\pm, *}$ correspond to the eigenstates of the pointer observable $Z_A$, and by construction automatically obey the Yanase condition.

It turns out that the measurement may be cast as an easily understood quantum circuit, in which the dilating system $\K$ constitutes three `register' qubits, initialized in the state $|001\>$. A von Neumann-L\"uders measurement would require that the states $|\pm\>_S\otimes|\Psi\>_\R \otimes |001\>_{1,2,3}$ evolve so that the eigenstate of the system is recorded in the computational basis of the register qubits, and can be read out by a measurement that respects the conservation constraint. The unitary over the composite system
 \bea\label{unitary}
V&=&(|0,0\>\<0,0|+|M,1\>\<M,1|)\otimes \I_{123}+\nonumber\\
&&\hspace{-2cm}\sum_{n=1}^{M}|\phi_{n}^-\>\<\phi_{n}^-|\otimes SWAP_{2,3}
+\sum_{n=1}^{M}|\phi_{n}^+\>\<\phi_{n}^+|\otimes SWAP_{1,3},
 \eea
and is represented as a quantum circuit in figure (\ref{circuit}).
The three register qubits simply correspond to ``+", ``-" and ``inconclusive". The projective measurement on the joint system is coupled to swap operations that shift the location of the ``1" in the register conditional on the outcome of the measurement. If the result of UD is ``+" it swaps register $2$ and $3$, if the result is ``-" it swaps register $1$ and $3$ and if the result is inconclusive it does nothing to the register qubits. By inspection, the unitary model corresponds to the optimal unambiguous discrimination protocol in the presence of the maximally asymmetric resource state $|\Psi\>$ for $\R$ and requires only a single unit of the conserved charge in the initial state of $\K$. This is optimal on the deterministic branch of $|\Psi\>$, as discussed either, and it can be shown that nothing is gained if we deviate off this branch, and so the performance is the optimal for such a scenario.

We might wonder if another discrimination criterion might be better satisfied by the above setup. For maximum likelihood estimation the analysis proceeds in a straightforward manner, and it turns out that the optimal POVM is achieved through the projective measurement $\{P_+=\sum_{n=}^M |\phi_n^+\>\<\phi_n^+|, P_-=\I -\sum_{n=1}^M |\phi_n^+\>\<\phi_n^+| \}$, which as a unitary circuit has the form
\bea\label{unitaryMLE}
V= P_+\otimes \I_{12}+P_-\otimes SWAP_{1,2}.
\eea
The probability of success is $\frac{M}{M+1}$, which is identical to the case of unambiguous discrimination. While $|\Psi\>$ is a maximally asymmetric state for our system, and so is optimal on its deterministic branch, the issue of its optimality overall is more subtle and is discussed in the next section where surprisingly we find that for the MLE criterion it is \emph{not} the optimal state over the full Hilbert space.

The issue of repeatability can also be simply understood within the quantum circuit example. If in addition to the resource state $|\Psi\>$ we also have another resource state $|\psi_{\mbox{\tiny copy}}\>=\frac{1}{2}(|0\>+|1\>)$ then we can simply adapt our quantum circuit so that the conditional swap gate also swaps in a fresh copy of $|e_+\>$ to $S$ in the event of a ``+" outcome, and in the event of a ``-" outcome performs a $\pi$-phase shift on $|\psi_{\mbox{\tiny copy}}\>$ and now swaps in a fresh copy of $|e_-\>$ to $S$. This ensures that in the event of a successful discrimination that the system $S$ is kept in its original eigenstate, however this does not provide perfect repeatability, since with some non-zero probability the discrimination stage will fail and so cannot algorithmically restore the system to its original state.

\subsection{Non-trivial asymmetry resources}\label{non-trivial-discrim}
In this section we give an example where the apparatus is of bounded-size, but the number of terms in its expansion is not bounded from above. Due to its practical importance we use a coherent state as our apparatus instead of a uniform superposition of number states. We compare the rate of increase of the probability of success in discriminating the two G-twirled states using maximum likelihood estimation (MLE) and unambiguous discrimination (UD).
\subsubsection{Unambiguous Discrimination}
We again consider the measurement of the observable $L_S$ with eigenstates $|+\>$ and $|-\>$, as in the previous section, but consider an asymmetry state of greater experimental relevance than the uniform superposition state $|\Psi\>$. Specifically, we use for our asymmetry resource system some infinite dimensional system $\R$, prepared in the zero phase coherent state $|\alpha\>=e^{-\frac{\alpha^2}{2}}\sum_{n=0}^{\infty}\frac{\alpha^n}{\sqrt{n!}}|n\>$, for which the states in the G-twirled ensemble are given by
\begin{eqnarray}
\bar{\rho}_\pm &=& \frac{1}{2}e^{-\alpha^2/2} |0,0\>\<0,0|+ \sum_{n \ge 1} \lambda_n |\phi_n^\pm\>\<\phi_n^\pm|.
\end{eqnarray}
We now have that the eigenstates of the G-twirled state in each sector are given by
\begin{eqnarray}
|\phi_n^\pm\>= \frac{\alpha}{\sqrt{\alpha^2 + n}} |n,0\> \pm \sqrt{\frac{n}{\alpha^2 +n}}|n-1,1\>,
\end{eqnarray}
with probability in the mixture given by
\begin{eqnarray}
\lambda_n = \frac{e^{-\alpha^2}\alpha^{2(n-1)}}{(n-1)!} \left(1+ \frac{\alpha^2}{n} \right ).
\end{eqnarray}

Again the two density operators are block diagonal in the eigenbasis of $N_{\mbox{\tiny tot}}$ and so our task again reduces to optimal discrimination within each sector, however this time the projected states are no longer orthogonal to each other. Within the sector $\H_n$ we have
\bea\label{projected g-twirled states}
\bar{\rho}_{\pm , n}&=&\frac{1}{1+\frac{\alpha^2}{n}}\left(
                                                  \begin{array}{cc}
                                                    \frac{\alpha^2}{n} & \pm\frac{\alpha}{\sqrt{n}} \\
                                                    \pm\frac{\alpha}{\sqrt{n}} & 1 \\
                                                  \end{array}
                                                \right),
\eea
each occurring with projection probability
\begin{eqnarray}
\Tr[\Pi_n\bar{\rho}_{\pm}]=\frac{1}{2}e^{-\alpha^2}\left (\frac{\alpha^{2n}}{n!}+\frac{\alpha^{2n-2}}{(n-1)!}\right ).
\end{eqnarray}

First we need to calculate the maximum probability of success in obtaining the conclusive result when trying to unambiguously discriminate the two projected states (\ref{projected g-twirled states}). These states have one dimensional kernels for which the problem of optimal UD admits a tidy solution \cite{Peres,Terry}. In order to satisfy the earlier condition (I), the discriminating POVM elements must be of the form $\{ a|\chi^+_n\>\<\chi^+_n|,b|\chi^-_n\>\<\chi^-_n|,\I-a|\chi^+_n\>\<\chi^+_n|-b|\chi^-_n\>\<\chi^-_n|\}$, where $|\chi^{\pm}_n\>\<\chi^{\pm}_n|$ are the projectors onto the kernels of $\rho^{\mp}_n$, and so the only variation parameters in the problem are the weights $a$ and $b$.

The optimal values of $a$ and $b$ are functions solely of the overlap probability of the two states $|\chi^\pm_n\>$ and the prior probabilities for the ensemble elements.

For our particular state the optimal POVM occurs for $a^{opt}=b^{opt}=\frac{2\alpha^2}{\alpha^2+n}$ which results in the maximum probability of success as
\bea
\sum_k p_{k,n}\Tr[\F_+(\bar{\rho}_{k,n})]=\left\{
                   \begin{array}{ll}
                     \frac{2n}{n+\alpha^2}, & \hbox{$n\leq \alpha^2$;} \\
                    \frac{2\alpha^2}{n+\alpha^2}, & \hbox{$n>\alpha^2$.}
                   \end{array}
                 \right.
\eea

Summing over all sectors we find that the optimal success probability to unambiguously discriminate the two states using the coherent state as the asymmetry resource is given by
\bea\label{PSUCC UD}
P_{\mbox{\tiny UD}}&=&1-e^{-\bar{N}}\frac{\bar{N}^{\bar{N}+1}}{\bar{N}!(1+\bar{N})},
\eea
where $\bar{N}=\<N\>$ is the expectation value of $N$ for the coherent state. In the large $\bar{N}$ limit, Stirling's approximation gives us that $P_{\mbox{\tiny UD}}\simeq1-\frac{1}{\sqrt{2\pi\bar{N}}}$.

\begin{figure}
\includegraphics[width=9.5cm]{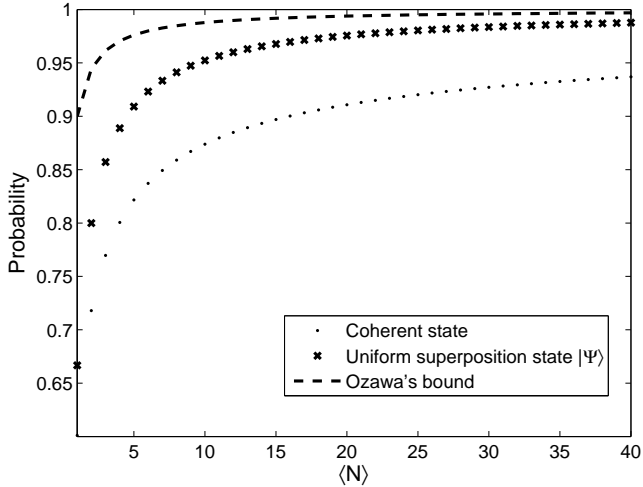}
\caption{Probability of success in unambiguous discrimination as a function of $\<N\>$ with a bound (dashed line) obtained by Ozawa of $1-(4+16\<N\>)^{-1}$, under a different criterion.}\label{UDuniform+coherent+OZAWA}
\end{figure}

\subsubsection{Maximum Likelihood Estimation}
We may alternatively, follow a maximum likelihood estimation route in which we compute the optimal discrimination, by once again restricting to the individual sectors. The probability of success in each sector $\H_n$ is $P_{\mbox{\tiny MLE,n}}=\frac{1}{2}\Tr[\bar{\rho}^{+}_{n}\Pi_{+,n}]+\frac{1}{2}\Tr[\bar{\rho}^{-}_{n}\Pi_{-,n}]$, where $\Pi_{+,n}$ and $\Pi_{-,n}$ are the POVM elements which we take to be projections and for simplicity we have taken $p_+=p_-=\frac{1}{2}$. This can be re-written as $P_{\mbox{\tiny MLE,n}}=\frac{1}{2}+\frac{1}{2}\Tr[(\bar{\rho}^+_n-\bar{\rho}^-_n)\Pi_{+,n}]$, which can be seen to take its maximum value
\bea
P_{\mbox{\tiny MLE,n}}=\frac{1}{2}+\left(\frac{\sqrt{n\bar{N}}}{n+\bar{N}}\right),
\eea
when $\Pi_{+,n}=\frac{1}{2}\left(
                        \begin{array}{cc}
                          1 & 1 \\
                          1 & 1 \\
                        \end{array}
                      \right)
$. Note that for a fixed value of $\bar{N}$ this probability increases from the value $\frac{1}{2}$ at $n=0$ to unit probability at $n=\bar{N}$ (if $\bar{N}\in \mathbb{N}$), before decreasing once more to $\frac{1}{2}$ as $n\rightarrow\infty$. Summing over the sectors, we find that the optimal probability for MLE on the G-twirled ensemble is
\bea\label{ProbMLE}
P_{\mbox{\tiny MLE}}=\frac{e^{-\bar{N}}}{4}\left[1+\sum_{n=1}^{\infty}\frac{\bar{N}^{n-1}}{(n-1)!}\left(1+\sqrt{\frac{\bar{N}}{n}}\right)^2\right].
\eea
\begin{figure}
\includegraphics[width=9.5cm]{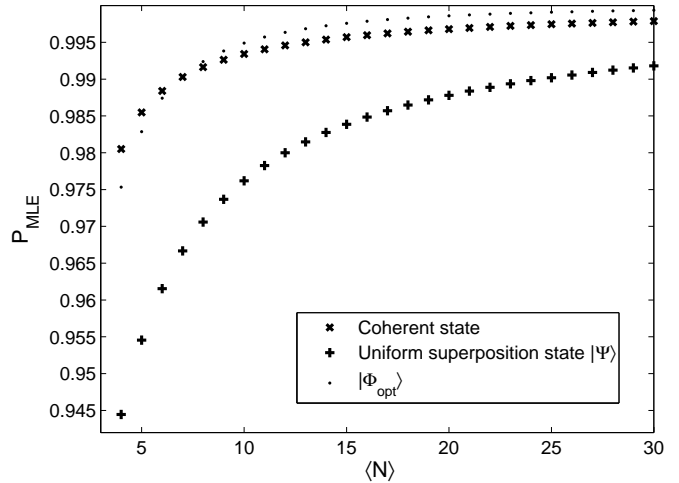}
\caption{Probability of MLE success as a function of $\<N\>$ for three different states: Coherent state, uniform superposition of number states and the optimal phase QRF.}\label{MLE3states}
\end{figure}
As can be seen from figure (\ref{MLE3states}), a bounded, infinite dimensional coherent state performs better than a finite uniform superposition state when we use the MLE as our criterion to discriminate between the states in the ensemble.

\subsection{The surprising case of optimality occurring off the deterministic branch of $|\Psi\>$.}

We have found that under the criterion of MLE discrimination it is possible to perform a probabilistic von-Neumann-L\"uders measurement with success probability $\frac{M}{M+1}$ for the uniform superposition state $|\Psi\>$, and with larger probability given by (\ref{ProbMLE}) in the case of a bounded coherent state. Any comparision between the coherent state and the state $|\Psi\>$ can rightly be questioned, given that the former is for an unbounded system and the dwindling amounts of asymmetry on its higher sectors $\H_{M+1}, \H_{M+2}, \cdots$, might be contributing enough to make the comparision unfair.

However the same cannot be said for states within the stochastic branch of $|\Psi\>$. As discussed, the state $|\Psi\>$ must be optimal on its deterministic branch, but it turns out that under the criterion of MLE it is not in fact the optimal state. In other words there exists a non-uniform state $|\Phi_{\mathrm{opt}}\>$ which has a higher probability of success, but where the stochastic conversion $|\Psi\> \longrightarrow |\Phi_{\mathrm{opt}}\>$ occurs with a sufficiently low probability that we are heavily penalised if we begin initially with the state $|\Psi\>$.

Maximum likelihood estimation was previously considered in the context of quantum phase reference frames \cite{Terrydegradation}, where a pure state quantum reference frame is used to distinguish between two states $|+\>$ and $|-\>$, as above. The analysis revealed that the optimal phase reference frame for a bounded system of dimension $M+1$ is given by
\begin{eqnarray}
|\Phi_{\mathrm{opt}}\>=C\sum_{n=0}^{M}\sin\left[\frac{(n+1)\pi}{M+2}\right]|n\>,
\end{eqnarray}
with the normalization constant $C$ is given by $C^{-2}=\frac{1}{4}(1+ 2M- \csc x  \sin [ (2M+1)x])+\sin^2[(M+1)x]$, and where $x=\frac{\pi}{M+2}$. This state provides us the globally optimal MLE success probability of
\bea
P_{\mbox{\tiny MLE}}=\frac{C^2}{2}\cos^2 \frac{x}{2} \left[M+2\cos x+\sin^2x\right]\nonumber
\eea
which, up to the order $1/M^2$ is given by $P_{\mbox{\tiny MLE}}\approx1-\frac{\pi^2}{4(M+2)^2}\approx 1-\frac{\pi^2}{16(\bar{N}+1)^2}$. In figure (\ref{MLE3states}), we find that $|\Phi_{\mathrm{opt}}\>$ does substantially better than $|\Psi\>$ and even outdoes the unbounded coherent state when used for MLE.

\section{Discussion}\label{discussion}

The WAY theorem, and related work, put fundamental limitations on the possible physical processes that quantum mechanics allows in the presence of a conservation law. Here we have reformulated this fundamental topic in terms of recent concepts coming from quantum information theory. In doing so, we have formulated a unified way of handling various scenarios that shed light on the origin of the fundamental constraints, provided a rigorous account of how optimal limits may be obtained under different criteria, and connected with the extensive literature on the theory of quantum state discrimination. We have also shown that any measuring apparatus $A$ naturally subdivides into a resource carrying component $\R$, and a readout component $\K$, that initially carries some sharp amount of conserved charge. The theory of resource asymmetry then provides us with the correct ordering of the set of all measuring apparatuses and also provides consistent measures for the accounting of internal resources. We have illustrated the subtleties that can arise, with the most asymmetric states not necessarily being optimal states for a given protocol. 

Within this viewpoint, the Yanase condition can now seen to be a statement that any readout measurement must fall within the resource theory constraints, and any measurements that do not obey this condition would imply some hidden asymmetry being smuggled into the accounting. 

One might take foundational issue with the very existence of any asymmetry resources in Nature, arguing that the full state of the universe must be symmetric under a particular symmetry group, and so worry that this forbids the types of measurement proceedures discussed in this paper. This turns out to not be an issue since it is perfectly consistent that the global state is symmetric, yet contains \emph{relational asymmetry}, where the reduced state on subsystems transforms non-trivially under the group action. This has previously been explored in the literature under the heading of protected (virtual) subsystems, both in the theory of quantum reference frames \cite{DJ,SSR-RF}, and the theory of robust, fault-tolerant protection for quantum information \cite{knillaflammeviola}. One could also simply pronounce that some superpositions (such as charge eigenstates \cite{ChargeSSR}) are fundamentally excluded by superselection rules, however, from the quantum reference frame perspective there is no essential difference between such an axiomatic prohibition of charge superposition and the statement that coherent superpositions for atom numbers are hard to prepare \cite{DowlingBartrudol}. All superpositions are prepared and defined relative to a particular reference frame, itself being a physical system, and any superselection rule can be taken as the empirical statement that we lack an appropriate reference frame state.

\begin{acknowledgments}
D.J. is supported by the Royal Commission for the Exhibition of 1851. M.A. and T.R. are supported by an EPSRC grant and the Leverhulme Trust. We would also like to thank Matt Palmer for helpful discussions.
\end{acknowledgments}


\begin{thebibliography}{99}
\bibitem{busch} P. Busch and M. Grabowski and P. J. Lahti, \textit{Operational Quantum Physics}, Springer (1997).
\bibitem{wigner} E. Wigner, Z. Phys. \textbf{133}, 101 (1952).
\bibitem{araki} H. Araki and M. M. Yanase, Phys. Rev. \textbf{120}, 622–626 (1960).
\bibitem{Yanase} M. M. Yanase, Phys. Rev. \textbf{123}, 666 (1961).
\bibitem{Ghirardi} G. C. Ghirardi, F. Miglietta, A. Rimini, and T. Weber, Phys. Rev. D \textbf{24}, 347 (1981)
\bibitem{Ozawa} M. Ozawa, Phys. Rev. Lett. \textbf{88}, 050402 (2002).
\bibitem{OzawaGUP} M. Ozawa, Phys. Rev. A \textbf{67}, 042105 (2003).
\bibitem{Leonbuschcontinuous} P. Busch, L. Loveridge, Phys. Rev. Lett. \textbf{106}, 110406 (2011).
\bibitem{Leonbuschreview} L. Loveridge and P. Busch, Eur. Phys. J. D \textbf{62}, 297 (2011).
\bibitem{ImanRob} I. Marvian, R. W. Spekkens, 	arXiv:1105.1816v1 (2011).
\bibitem{ImanRobreview} I. Marvian, R. W. Spekkens, arXiv:1104.0018v1 (2011).
\bibitem{Nielsen&Chuang} M. A. Nielsen and I. L. Chuang, \textit{Quantum Computation and Quantum Information} (Cambridge University Press, Cambridge, England, 2000).
\bibitem{DJ} D. Jennings, Phys. Rev. A \textbf{84}, 012306 (2011).
\bibitem{vanEnk} S.J. van Enk, Phys. Rev. A \textbf{71}, 032339 (2005).
\bibitem{SSR-RF} S. D. Bartlett, T. Rudolph and R. W. Spekkens, Rev. Mod. Phys. \textbf{79}, 555–609 (2007).
\bibitem{AJR} M. Ahmadi, D. Jennings, and T. Rudolph, Phys. Rev. A \textbf{82}, 032320 (2010).
\bibitem{GourSpek} G. Gour and R. W. Spekkens, New J. Phys. \textbf{10}, 033023 (2008).
\bibitem{ImanGiladRobREF} G. Gour, I. Marvian  and R. W. Spekkens, Phys. Rev. A \textbf{80}, 012307 (2009).
\bibitem{KeylWerner} M. Keyl and R. F. Werner, J. Math. Phys. \textbf{40}, 3283 (1999).

\bibitem{SJ1} S. Jevtic, D. Jennings, T. Rudolph, Phys. Rev. Lett. \textbf{108} 110403 (2012).
\bibitem{SJ2} S. Jevtic, D. Jennings, T. Rudolph, Phys. Rev. A \textbf{85} 052121 (2012).
\bibitem{steveterryrobQC} S. Bartlett, T. Rudolph, R. Spekkens, and P. Turner, New J. Phys. \textbf{11}, 063013 (2009).
\bibitem{Raynal} P. Raynal, PhD thesis, arXiv:quant-ph/0611133v1 (2006).
\bibitem{Peres} A. Peres, Phys. Lett. A \textbf{128}, 19 (1988).
\bibitem{Terry} T. Rudolph, R. W. Spekkens and P. S. Turner, Phys. Rev. A \textbf{68}, 010301(R) (2003).
\bibitem{Terrydegradation} S. Bartlett, T. Rudolph, R. Spekkens, and P. S. Turner, New J. Phys. \textbf{8}, 58 (2006).
\bibitem{knillaflammeviola} E. Knill, R. Laflamme, and L. Viola, Phys. Rev. Lett. \textbf{84}, 2525 (2000).
\bibitem{ChargeSSR} Y. Aharonov and L. Susskind, Phys. Rev. \textbf{155}, 1428–1431 (1967).
\bibitem{DowlingBartrudol} M. R. Dowling, S. D. Bartlett, T. Rudolph, and R. W. Spekkens, Phys. Rev. A \textbf{74}, 052113 (2006).

\end{thebibliography}
\end{document}